\documentclass[12pt]{iopart}
\usepackage{iopams}
\usepackage[T1]{fontenc}

\usepackage{graphicx}

\begin{document}
\title{Reconstruction of scalar potentials in two-field cosmological models}

\author{Alexander A. Andrianov}
\address{
V.A. Fock Department of Theoretical Physics, Saint Petersburg State University,
198904, S.-Petersburg, Russia \\
Departament d'Estructura i Constituents de la Materia,
Universitat de Barcelona, 08028, Barcelona, Spain}
\author{Francesco Cannata}
\address{INFN, Via Irnerio 46, 40126 Bologna,
Italy}
\author{Alexander Y. Kamenshchik}
\address{Dipartimento di Fisica and INFN, Via Irnerio 46, 40126 Bologna,
Italy\\
L.D. Landau Institute for Theoretical Physics of the Russian
Academy of Sciences, Kosygin str. 2, 119334 Moscow, Russia }
\author{Daniele Regoli}
\address{Dipartimento di Fisica, Via Irnerio 46, 40126 Bologna,
Italy}
\eads{\mailto{andrianov@bo.infn.it}, \mailto{cannata@bo.infn.it}, \mailto{kamenshchik@bo.infn.it} and \mailto{astapovo@gmail.com}}
\begin{abstract}
We study the procedure of the reconstruction of phantom-scalar field potentials in two-field cosmological models.
It is shown that while in the one-field case the chosen cosmological evolution defines uniquely the form of the scalar
potential, in the two-field case one has an infinite number of possibilities. The classification of a large class of possible
potentials is presented and the dependence of  cosmological dynamics on the choice of initial conditions is
investigated qualitatively and numerically for two particular models. 
\end{abstract}

\hfill Keywords: phantom dark energy, scalar fields\\

\submitto{JCAP}
\maketitle

\section{Introduction}
The discovery of cosmic acceleration \cite{cosmic} has
stimulated the construction of a class of dark energy models
\cite{dark,quintessence,darkmodel} describing this effect.
Notice that the cosmological models based on  scalar fields were considered
long before the observational discovery of cosmic acceleration \cite{before}.
The dark energy should possess a negative pressure such that the relation
between pressure and energy density $w$ is less than $-1/3$.
The present day value of the parameter $w$ is close to $-1$ (cosmological constant) but some observations indicate that the 
value of this parameter slightly less than $-1$ provides the best fit.
The corresponding dark energy has been named phantom dark
energy \cite{phantom}.
According to some authors, the analysis of observations
permits the existence of the moment when the universe
changes the value of the parameter $w$ from $w > -1$ to
$w < -1$ \cite{phant-obs,phant-obs1}. This transition is
called ``the crossing of the phantom divide line''.
The most recent investigations  have shown that
the phantom divide line crossing is still not excluded by the
data \cite{obs-new}.

It is easy to see, that the standard minimally coupled scalar
field cannot give rise to the phantom dark energy, because in this model
the absolute value of  energy density is always greater than that of
pressure, i.e. $|w| < 1$. A possible  way out of this situation
is the consideration of the scalar field models with the negative
kinetic term.
 Thus, the important problem arising in connection with
the phantom energy is  the crossing of the phantom
divide line. The general belief is that while this crossing is not admissible
in simple minimally coupled models its  explanation
 requires  more complicated models such as ``multifield'' ones
or models with non-minimal coupling between scalar field and gravity
(see e.g. \cite{divide}).

In preceding papers some of us \cite{we,we1}
described the phenomenon of the change
of sign of the kinetic term of the scalar field implied by the Einstein
equations. It was shown that such a change is possible only when the potential of the
scalar field possesses some cusps and, moreover, for some very special
initial conditions on the time derivatives and values of the
considered scalar field approaching to the phantom divide line.
At the same time,
two-field models  including one standard scalar field and
a phantom field can describe the phenomenon of
the (de)-phantomization under very general conditions and using rather
simple potentials \cite{two-field,two-field1}.
In the present paper we would like to attract  attention to a drastic
difference between two- and one-field models.

The reconstruction procedure of the scalar field potential models
is well-known \cite{Star,Barrow,we-tach,Yurov,Ohta,Zhuk,Szydlowski,Vernov}. We recapitulate it briefly. The cosmological evolution in a
flat Friedmann model with the metric
\begin{equation}
ds^2 = dt^2- a^2(t)dl^2
\label{Fried}
\end{equation}
can be given by the time evolution of the Hubble parameter $h(t) \equiv \frac{\dot{a}}{a}$, which
satisfies the Friedmann equation
\begin{equation}
h^2 = \varepsilon,
\label{Fried1}
\end{equation}
where $\varepsilon$ is the energy density and a convenient normalization of the Newton constant is chosen.
Differentiating equation (\ref{Fried1}) and using the energy conservation equation
\begin{equation}
\dot{\varepsilon} = -3h(\varepsilon + p),
\label{en-conserv}
\end{equation}
where $p$ is the pressure, one comes to
\begin{equation}
\dot{h} = -\frac32 (\varepsilon + p).
\label{h-dot}
\end{equation}
If the matter is represented by a spatially homogeneous minimally coupled scalar field, then
the energy density and the pressure are given by the formulae
\begin{equation}
\varepsilon = \frac12\dot{\phi}^2 + V(\phi),
\label{energy}
\end{equation}
\begin{equation}
p = \frac12\dot{\phi}^2 - V(\phi),
\label{pressure}
\end{equation}
where $V(\phi)$ is a scalar field potential.
Combining equations (\ref{Fried1}), (\ref{h-dot}), (\ref{energy}), (\ref{pressure})
we have
\begin{equation}
V = \frac{\dot{h}}{3} + h^2,
\label{poten}
\end{equation}
and
\begin{equation}
\dot{\phi}^2 = -\frac23 \dot{h}.
\label{phi-dot}
\end{equation}
Equation (\ref{poten}) represents the potential as a function of time $t$. Integrating equation (\ref{phi-dot})
one can find the scalar field as a function of time. Inverting this dependence we can obtain the time parameter as
a function of $\phi$ and substituting the corresponding formula into equation (\ref{poten}) one arrives to the uniquely
reconstructed potential $V(\phi)$. It is necessary to stress that this potential reproduces a given cosmological evolution
only for some special choice of initial conditions on the scalar field and its time derivative \cite{we-tach,we}.
Below we shall show that in the case of two scalar fields one has an enormous freedom in the choice of
the two-field potential providing the same cosmological evolution. This freedom is connected with the fact
that the kinetic term has now two contributions. In order to examine the problem of phantom divide line crossing we shall be interested here in the case of one standard
scalar field and one phantom field $\xi$ ,
whose kinetic term has a negative sign. In this case the total energy density and pressure will be given by
\begin{equation}
\varepsilon = \frac12\dot{\phi}^2 - \frac12\dot{\xi}^2 + V(\phi,\xi),
\label{energy1}
\end{equation}
\begin{equation}
p = \frac12\dot{\phi}^2 - \frac12\dot{\xi}^2 - V(\phi,\xi).
\label{pressure1}
\end{equation}
The relation (\ref{poten}) expressing the potential as a function of $t$ does not change in form, but instead  of
equation (\ref{phi-dot}) we have
\begin{equation}
\dot{\phi}^2 - \dot{\xi}^2  = -\frac23 \dot{h}.
\label{phi-dot1}
\end{equation}
Now, one has rather a wide freedom in the choice of the time dependence of one of two fields. After that the time dependence
of the second field can be found from equation (\ref{phi-dot1}). However, the freedom is not yet exhausted. Indeed, having two representations for the time parameter $t$ as a function of $\phi$ or $\xi$, one can construct an infinite number of potentials
$V(\phi,\xi)$ using the formula (\ref{poten}) and some rather loose consistency conditions. It is rather difficult to characterize
all the family of possible two-field potentials, reproducing given cosmological evolution $h(t)$. In the present paper, we describe
some general principles of construction of such potentials, then we consider some concrete examples.

\section{The system of equations for two-field cosmological model}
The system of equations, which we  study contains (\ref{poten}) and (\ref{phi-dot1}) and two Klein-Gordon equations
\begin{equation}
\ddot{\phi} + 3h\dot{\phi} + \frac{\partial V(\phi,\xi)}{\partial \phi} = 0,
\label{KG}
\end{equation}
\begin{equation}
\ddot{\xi} + 3h\dot{\xi} - \frac{\partial V(\phi,\xi)}{\partial \xi} = 0.
\label{KG1}
\end{equation}
From equations (\ref{KG}) and (\ref{KG1}) we can find the partial derivatives $\frac{\partial V(\phi,\xi)}{\partial \phi}$
and $\frac{\partial V(\phi,\xi)}{\partial \xi}$ as functions of time $t$. The consistency relation
\begin{equation}
\dot{V} = \frac{\partial V(\phi,\xi)}{\partial \phi}(t)\dot{\phi} + \frac{\partial V(\phi,\xi)}{\partial \xi}(t)\dot{\xi}
\label{consistency}
\end{equation}
is respected.

Before starting the construction of potentials for particular cosmological evolutions, it it useful to
consider some  mathematical aspects of the problem of reconstruction of a function of two variables in general terms.

\subsection{Reconstruction of the function of two variables, which in turn depends on a third parameter}
Let us consider the function of two variables $F(x,y)$  defined on a curve, parameterized by $t$. Suppose that we know
the function $F(t)$ and its partial derivatives  as functions of $t$:
\begin{equation}
F(x(t),y(t)) = F(t),
\label{function}
\end{equation}
\begin{equation}
\frac{\partial F(x(t),y(t))}{ \partial x} = \frac{\partial F}{ \partial x}(t),
\label{partial}
\end{equation}
\begin{equation}
\frac{\partial F(x(t),y(t))}{ \partial y} = \frac{\partial F}{ \partial y}(t).
\label{partial1}
\end{equation}
These three functions should satisfy the consistency relation
\begin{equation}
\dot{F}(t) = \frac{\partial F}{ \partial x}(t)\dot{x} + \frac{\partial F}{ \partial y}(t)\dot{y}.
\label{consistency1}
\end{equation}

As a simple example we can consider the curve
\begin{equation}
x(t) = t,\ \ y(t) = t^2,
\label{curve}
\end{equation}
while
\begin{equation}
F(t) = t^2,
\label{example}
\end{equation}
\begin{equation}
\frac{\partial F}{\partial x} = t^2,
\label{example1}
\end{equation}
\begin{equation}
\frac{\partial F}{\partial y} = t,
\label{example2}
\end{equation}
and equation (\ref{consistency1}) is satisfied.

Thus, we would like to reconstruct the function $F(x,y)$ having explicit expressions in right-hand side of equations (\ref{function})--
(\ref{partial1}). This reconstruction is not unique. We shall begin the reconstruction process taking such simple ansatzes
as
\begin{equation}
F_1(x,y) = G_1(x) + H_1(y),
\label{sum}
\end{equation}
\begin{equation}
F_2(x,y) = G_2(x)H_2(y),
\label{product}
\end{equation}
\begin{equation}
F_3(x,y) = (G_3(x) + H_3(y))^{\alpha}.
\label{power}
\end{equation}

The assumption (\ref{sum})  immediately implies
\begin{equation}
\frac{\partial F_1}{\partial x} = \frac{\partial G_1}{\partial x},
\label{partial2}
\end{equation}
\begin{equation}
\frac{\partial F_1}{\partial y} = \frac{\partial H_1}{\partial y}.
\label{partial3}
\end{equation}
Therefrom  one obtains
\begin{equation}
G_1(x) = \int^x \frac{\partial F_1}{\partial x'}(t(x')) dx',
\label{G}
\end{equation}
\begin{equation}
H_1(y) = \int^y \frac{\partial H_1}{\partial y'}(t(y')) dy'.
\label{H}
\end{equation}
Hence
\begin{equation}
F_1(x,y) = \int^x \frac{\partial F_1}{\partial x'}(t(x')) dx' + \int^y \frac{\partial H_1}{\partial y'}(t(y')) dy'.
\label{result}
\end{equation}
For an example given by equations (\ref{curve})--(\ref{example2}) the function $F_1(x,y)$ is
\begin{equation}
F_1(x,y) = \int^x t^2(x')dx' + \int^y t(y')dy' = \int^x x'^2 dx' \pm \int^y \sqrt{y'} dy'.
\label{example3}
\end{equation}
Explicitly
\begin{equation}
F_1(x,y)=
\cases{
\frac{1}{3}\left( x^3+2y^{3/2}\right),& x>0,\\
\frac{1}{3}\left( x^3-2y^{3/2}\right),& x<0.}\label{e:ex_somma}
\end{equation}

Similar reasonings give for the assumptions (\ref{product}),and  (\ref{power})  correspondingly
\begin{equation}
F_2(x,y) = \exp\left\{\int \left(\frac1F\frac{\partial F}{\partial x}\right)(t(x))dx
+ \int \left(\frac1F\frac{\partial F}{\partial y}\right)(t(y))dy\right\},
\label{F2}
\end{equation}
\begin{equation}
F_3(x,y) = \left\{\int \frac{dx}{\alpha} \left(F^{\frac{1}{\alpha}-1}\frac{\partial F}{\partial x}\right)(t(x))
+ \int \frac{dy}{\alpha} \left(F^{\frac{1}{\alpha}-1}\frac{\partial F}{\partial y}\right)(t(y))\right\}^{\alpha}.
\label{F3}
\end{equation}
For our simple example (\ref{curve})--(\ref{example2}) the functions $F_2(x,y)$ and $F_3(x,y)$ have the form
\begin{equation}
F_2(x,y) = xy,
\label{exampleF2}
\end{equation}
\begin{equation}
F_3(x,y)=
\cases{
\left[\frac{1}{3}\left(x^{3/\alpha} +2y^{3/2\alpha}\right)\right]^{\alpha}, & x>0,\\
\left[\frac{1}{3}\left(x^{3/\alpha}(-)^{3/\alpha}2y^{3/2\alpha}\right)\right]^{\alpha}, & x<0.}
\end{equation}
Thus, we have seen that the same input of ``time'' functions (\ref{example})--(\ref{example2}) on the curve (\ref{curve})
produces quite different functions of variables $x$ and $y$.

Naturally, one can introduce many other assumptions for reconstruction of $F(x,y)$. For example, one can consider
linear combinations of $x$ and $y$ as functions of the parameter $t$ and decompose the presumed function $F$ as a sum
or a product of the functions of these new variables.

Now we present a way of constructing  the whole family of solutions starting from a given one. Let us suppose that we have
a function $F_0(x,y)$ satisfying all the necessary conditions. Let us take an arbitrary function
\begin{equation}
f(x,y) = f\left(\frac{t(x)}{t(y)}\right),
\label{f-small}
\end{equation}
which depends only on the ratio $t(x)/t(y)$. We require also
\begin{equation}
f(1) = 1,\ \ f'(1) = 0,
\label{f-small1}
\end{equation}
i.e. the function reduces to unity and its derivative vanishes on the curve $(x(t),y(t))$.
Then it is obvious that
the function
\begin{equation}
F(x,y) = F_0(x,y)f\left(\frac{t(x)}{t(y)}\right)
\label{F-new}
\end{equation}
is also a solution. This permits us to generate a whole family of solutions, depending on a choice of the
function $f$.
Moreover, one can construct other solutions, adding to the function $F(x,y)$ a term proportional to
$(t(x)-t(y))^2$.

\subsection{Cosmological applications, an evolution ``Bang to Rip''}
To show how this procedure works in cosmology, we consider a relatively simple cosmological evolution, which
nevertheless is of  particular interest, because it describes the phantom divide line crossing.
Let us suppose that the Hubble variable for this evolution behaves as
\begin{equation}
h(t) = \frac{A}{t(t_R - t)},
\label{Hubble-bang-to-rip}
\end{equation}
 where $A$ is a positive constant. At the beginning of the cosmological evolution, when $t \rightarrow 0$ the universe
is born from the standard Big Bang type cosmological singularity, because $h(t) \sim 1/t$. Then, when $t \rightarrow t_R$,
the universe is superaccelerated, approaching the Big Rip singularity $h(t) \sim 1/(t_R - t)$.
Substituting the function (\ref{Hubble-bang-to-rip}) and its time derivative into equations. (\ref{phi-dot1}) and (\ref{poten})
we come to
\begin{equation}
\dot{\phi}^2 - \dot{\xi}^2 = - \frac{2A(2t-t_R)}{3t^2(t_R-t)^2},
\label{BtR}
\end{equation}
\begin{equation}
V(t) = \frac{A(2t-t_R+3A)}{3t^2(t_R-t)^2}.
\label{poten1}
\end{equation}
For convenience we
choose also the parameter $A$ as
\begin{equation}
A = \frac{t_R}{3},
\label{special-choice}
\end{equation}
Then,
\begin{equation}
h(t) = \frac{t_R}{3t(t_R-t)},
\label{Hubble-simple}
\end{equation}
and
\begin{equation}
V(t) = \frac{2t_R}{9t(t_R-t)^2}.
\label{poten2}
\end{equation}

Let us consider now a  special choice of functions $\phi(t)$ and $\xi(t)$ used already\footnote{Notice that the origin of two scalar fields has been associated in \cite{two-field} with a non-Hermitian complex scalar field theory and there a classical solution was found as a saddle point in  "double" complexification.} in \cite{two-field},
\begin{equation}
\phi(t) = -\frac43 {\rm arctanh} \sqrt{\frac{t_R-t}{t_R}},
\label{phi1}
\end{equation}
\begin{equation}
\xi(t) = \frac{\sqrt{2}}{3} \ln \frac{t}{t_R-t}.
\label{xi1}
\end{equation}
The derivatives of the potential with respect to the fields $\phi$ and $\xi$ could be found from the Klein-Gordon equations
(\ref{KG}) and (\ref{KG1}):
\begin{equation}
\frac{\partial V}{\partial \xi} = \ddot{\xi} + 3h\dot{\xi} = \frac{\sqrt{2}}{3}\frac{2t_R}{t(t_R-t)^2},
\label{der-pot}
\end{equation}
\begin{equation}
\frac{\partial V}{\partial \phi} = -\ddot{\phi} - 3h\dot{\phi} = -\frac{\sqrt{t_R}}{t(t_R-t)^{3/2}}.
\label{der-pot1}
\end{equation}
We can obtain also the time parameter as a function of $\phi$ or $\xi$:
\begin{equation}
t(\phi) = \frac{t_R}{\cosh^2(-3\phi/4)},
\label{time-phi}
\end{equation}
\begin{equation}
t(\xi) = \frac{t_R}{\exp(-3\xi/\sqrt{2})+1}.
\label{time-xi}
\end{equation}

Now we can make a hypothesis about the structure of the potential $V(\xi,\phi)$:
\begin{equation}
V_2(\xi,\phi) = G(\xi)H(\phi).
\label{hypo}
\end{equation}
Applying the technique described in the subsection A, we can get $G(\xi)$:
\begin{eqnarray}
&&\ln G(\xi) = \int \left(\frac1V \frac{\partial V}{\partial \xi}\right)(t(\xi)) d\xi  \nonumber \\
&&= \int \frac{9t(\xi)(t(\xi)-t_R)^2}{2t_R}\frac{\sqrt{2}}{3}\frac{2t_R}{t(\xi)(t_R-t(\xi))^2}d\xi = 3\sqrt{2}\xi,
\label{G-hypo}
\end{eqnarray}
 and
\begin{equation}
G(\xi) = \exp(3\sqrt{2}\xi).
\label{G-hypo1}
\end{equation}
To find $H(\phi)$ one can use the analogous direct integration, but we prefer to implement a formula
\begin{equation}
H(\phi) = \frac{V(t(\phi)}{G(\xi(t(\phi)))},
\label{another}
\end{equation}
which gives
\begin{equation}
H(\phi(t)) = \frac{2t_R}{9t^3},
\label{H-new}
\end{equation}
and hence,
\begin{equation}
H(\phi) = \frac{2}{9t_R^2}\cosh^6(-3\phi/4).
\label{H-new1}
\end{equation}
Finally,
\begin{equation}
V_2(\xi,\phi) = \frac{2}{9t_R^2}\cosh^6(-3\phi/4)\exp(3\sqrt{2}\xi).
\label{V-2new}
\end{equation}
Here we have reproduced the potential studied in \cite{two-field}.

Making the choice
\begin{equation}
V_1(\xi,\phi) = G(\xi) + H(\phi),
\label{choice1}
\end{equation}
we derive
\begin{eqnarray}
V_{1}(\xi,\phi) = \frac{2}{3t_R^2}\left[-\frac13e^{-3\xi/\sqrt{2}} + 3\sqrt{2}\xi
+2e^{3\xi/\sqrt{2}}
+ \frac{1}{3} e^{6\xi/\sqrt{2}} \right. \nonumber \\
\left.+\frac{\sinh^4(-3\phi/4) + \sinh^2(-3\phi/4) -1}{\sinh^2(-3\phi/4)}
+ \ln \sinh^4(-3\phi/4)\right].
\label{poten3}
\end{eqnarray}

Now, we can make another choice of the field functions $\phi(t)$ and $\xi(t)$, satisfying  the condition
(\ref{BtR}):
\begin{equation}
\phi(t) = \frac{\sqrt{2}}{3} \ln \frac{t}{t_R-t},
\label{phi-new3}
\end{equation}
\begin{equation}
\xi(t) = \frac{4}{3} {\rm arctanh} \sqrt{\frac{t}{t_R}}.
\label{xi-new3}
\end{equation}
The time parameter $t$ is a function of fields is
\begin{equation}
t(\phi) = \frac{t_R}{\exp(-3\phi/\sqrt{2})+1},
\label{time-phi1}
\end{equation}
\begin{equation}
t(\xi) = \frac{t_R}{\tanh^2(3\xi/4)}.
\label{time-xi1}
\end{equation}

Looking for the potential as a sum of functions of two fields as in equation (\ref{choice1}) after lengthy but
straightforward calculations we come to the following potential:
\begin{eqnarray}
V_1(\xi,\phi) = &\frac{2}{3t_R^2}\left[1 + \frac23 \sinh^4 (3\xi/4) + 3\sinh^2 (3\xi/4)
- \frac{1}{3\sinh^2 (3\xi/4)} \right. \nonumber \\
&+ 2 \ln \sinh^2 (3\xi/4) + \frac23 \exp (-3\phi/\sqrt{2}) - 3\sqrt{2}\phi \nonumber \\
&\left. - 2\exp(3\phi/\sqrt{2})-\frac13\exp(\phi/\sqrt{2})\right].
\label{poten5}
\end{eqnarray}

Similarly for the potential designed as a product of functions of two fields (\ref{hypo})
we obtain
\begin{equation}
V_2(\xi,\phi) = \frac{2}{9t_R^2}\sinh^2(3\xi/4)\cosh^2(3\xi/4)\exp(-3\sqrt{2}\phi).
\label{poten6}
\end{equation}

One can make also other choices of functions $\phi(t)$ and $\xi(t)$ generating other potentials, but we shall not do it here
(see \cite{Regoli}),  concentrating instead on the qualitative and numerical analysis of two toy cosmological models described by
potentials (\ref{poten5}) and (\ref{V-2new}).

\section{Analysis of cosmological models}
It is well known \cite{Bel-Khal}  that for the qualitative analysis of the system of cosmological
equations it is convenient to present it as a dynamical system, i.e. a system of first-order differential equations.
Introducing the new variables $x$ and $y$ we can write
\begin{equation}\left\{
\begin{array}{ll}
\dot{\phi} &= x,  \\
\dot{\xi} &= y,  \\
\dot{x}& =-3{\rm sign}(h)x \sqrt{\frac{x^2}{2} - \frac{y^2}{2} + V(\xi,\phi)} - \frac{\partial V}{\partial \phi},
\\
\dot{y} &=-3{\rm sign}(h)y \sqrt{\frac{x^2}{2} - \frac{y^2}{2} + V(\xi,\phi)} - \frac{\partial V}{\partial \xi}.

\end{array}\right.\label{system}
\end{equation}
Notice that the reflection
\begin{eqnarray}
&&x \rightarrow -x, \nonumber \\
&&y \rightarrow -y, \nonumber \\
&&t \rightarrow -t
\label{inversion}
\end{eqnarray}
transforms the system into one describing the cosmological evolution with the opposite sign of the Hubble parameter.
The stationary points of the system (\ref{system}) are given by
\begin{equation}
x = 0,\ y = 0,\ \frac{\partial V}{\partial \phi} = 0,\ \frac{\partial V}{\partial \xi} = 0.
\label{stationary}
\end{equation}

\subsection{Model I}
In this subsection we shall analyze the cosmological model with two fields - standard scalar and phantom, described by
the potential (\ref{poten5}).
For this potential the system of equations (\ref{system})  reads

\begin{equation}\left\{
\begin{array}{ll}
\dot\phi=&x,\\
\dot\xi=&y,\\
\dot x=&-3{\rm sign}(h)x\sqrt{\frac{x^2}{2}-\frac{y^2}{2}+\frac{2}{9t_R^2}\sinh^2(3\xi/4)\cosh^6(3\xi/4)e^{-3\sqrt{2}\phi}} \\
&+\frac{2\sqrt{2}}{3t_R^2}\sinh^2(3\xi/4)\cosh^6(3\xi/4)e^{-3\sqrt{2}\phi}, \\ 
\dot y=&-3{\rm sign}(h)y\sqrt{\frac{x^2}{2}-\frac{y^2}{2}+\frac{2}{9t_R^2}\sinh^2(3\xi/4)\cosh^6(3\xi/4)e^{-3\sqrt{2}\phi}} \\&+\frac{\sinh(3\xi/4)\cosh^5(3\xi/4)}{3t_R^2}\left[\frac{1}{3}\cosh^2(3\xi/4)+\sinh^2(3\xi/4)\right]e^{-3\sqrt{2}\phi}.
\end{array}\right.\label{e:sistV4}
\end{equation}

It is easy to see that there are stationary points
\begin{equation}
\phi = \phi_0,\  \xi = 0,\  x = 0,\  y = 0,
\label{stat}
\end{equation}
where $\phi_0$ is arbitrary.
For these points the potential and hence the Hubble variable vanish. Thus, we have a static cosmological solution.
We should study the behavior of our system in the neighborhood of the point (\ref{stat}) in linear approximation:
\begin{equation}\left\{
\begin{array}{ll}
\dot\phi=&x,\\
\dot\xi=&y,\\
\dot x=&0\\
\dot y=&+\frac{\xi}{4t_R^2}e^{-3\sqrt{2}\phi_0}.
\end{array}\right.\label{e:sistV4_li}
\end{equation}

One sees that the dynamics of $\phi$ in this approximation is frozen and hence we can focus on the study
of the dynamics of the variables $\xi,y$. The eigenvalues of the corresponding subsystem of two equations
are
\begin{equation}
\lambda_{1,2} = \mp \frac{e^{-3\phi_0/\sqrt{2}}}{2t_R}.
\label{saddle}
\end{equation}
These eigenvalues are real and have opposite signs, so one has a saddle point in the plane $(\xi,y)$ and
this means that the points (\ref{stat}) are  unstable.

One can make another qualitative observation. Freezing the dynamics of $\xi$ independently of $\phi$,
namely choosing $y = 0, \xi =0$, which implies also $\dot{y} = \ddot{\xi} = 0$, one has the following equation of motion for
$\phi$:
\begin{equation}
\ddot{\phi} + 3h\phi = 0.
\label{massless}
\end{equation}
Equation (\ref{massless}) is nothing but the Klein-Gordon equation for a massless scalar field on the Friedmann background,
whose solution is
\begin{equation}
\phi(t) = \frac{\sqrt{2}}{3}\ln t,
\end{equation}
and which gives a Hubble variable
\begin{equation}
h(t) = \frac{1}{3t}.
\end{equation}
This is an evolution of the flat Friedmann universe, filled with stiff matter with the equation of state $p =\varepsilon$.
It describes a universe, born from the Big Bang singularity and infinitely expanding. Naturally, for the opposite sign
of the Hubble parameter, one has the contracting universe ending in the Big Crunch cosmological singularity.

\begin{figure}[htp]\centering
\includegraphics{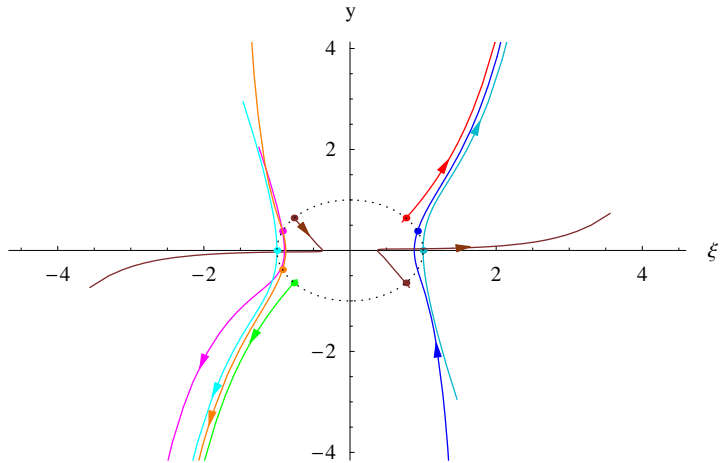}
\caption{An example of section of the 4D phase space obtained with numerical calculations. We can see a series of trajectories corresponding to different choices of the initial conditions. Every initial condition in the graphic is chosen on the dashed ``ellipse'' centered in the origin and is emphasized by a colored dot. A ``saddle point-like'' structure of the set  of the trajectories clearly emerges.}\label{figure1}
\end{figure}
\begin{figure}[htp]\centering
\includegraphics[scale=.7]{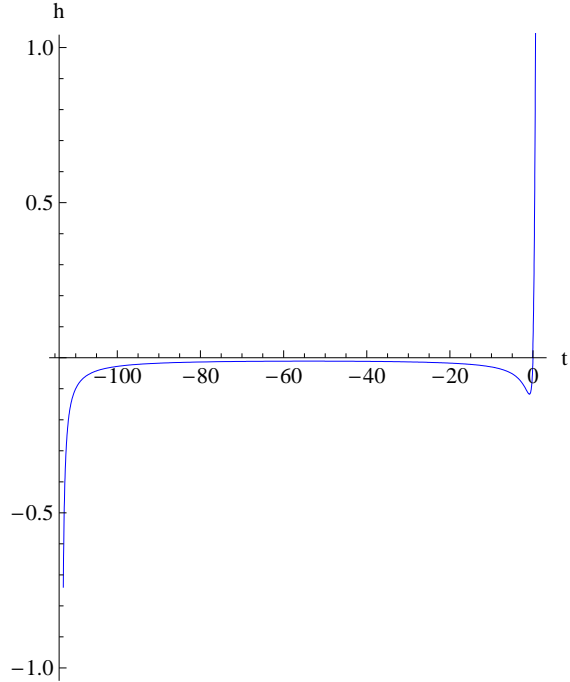}
\caption{Typical behavior of the Hubble parameter for the model I trajectories. The presence of two stationary points (namely a maximum followed by a minimum) indicates the double crossing of the phantom divide line. Both the initial and final singularities are  characterized by a Big Rip behavior, the first is a  contraction and the second is an expansion.}\label{figure2}
\end{figure}

Now, we describe some results of numerical calculations to have an idea about the structure of the set 
of possible cosmological evolutions coexisting in the model under consideration. 
We have carried out two kinds of simulations. First, we have considered neighborhood of the plane 
$y,\xi$ with the initial conditions on the field $\phi$ such that $\phi(0) = 0, \dot\phi (0) = 0$ (see Figure \ref{figure1}). The initial conditions 
for the phantom field were chosen in such a way that the sum of absolute values of the kinetic and potential energies were 
fixed. Then, running the time back and forward we have seen that the absolute majority of the cosmological evolutions began 
at the singularity of the ``anti-Big Rip'' type (Figure \ref{figure2}). Namely, the initial cosmological singularity were characterized by an 
infinite value of the cosmological radius and an infinite negative value of its time derivative (and also of the Hubble variable).
Then the universe squeezes, being dominated by the phantom scalar field $\xi$. At some moment the universe passes the phantom divide line and the universe continues squeezing but with $\dot{h} < 0$. Then it achieves the minimal value of the cosmological radius 
and an expansion begins. At some moment the universe undergoes the second phantom divide line crossing and its expansion becomes
super-accelerated culminating in an encounter with a Big Rip singularity. Apparently this scenario is very different from the standard cosmological scenario and with its phantom version Bang-to-Rip, which has played a role of an input in the construction 
of our potentials.
The second procedure, which we have used is the consideration of trajectories close to our initial trajectory of the Bang-to-Rip type. The numerical analysis shows that this trajectory is unstable and the neighboring trajectories again have anti-Big Rip -
double crossing of the phantom line - Big Rip behavior described above.  
However, it is necessary to emphasize that a small subset of the trajectories of the Bang-to-Rip type exist, being not 
in the vicinity of our initial trajectory.

\subsection{Model II}
In this subsection we shall study the cosmological model with the potential (\ref{V-2new}). Now the system of equations
(\ref{system}) looks like
\begin{equation}\left\{
\begin{array}{ll}
\dot\phi=&x,\\
\dot\xi=&y, \\
\dot x=&-3{\rm sign}(h)x\sqrt{\frac{x^2}{2}-\frac{y^2}{2}+\frac{2}{9t_R^2}\cosh^6(3\phi/4) \exp\{3\sqrt{2}\xi\}} \\
&-\frac{1}{t_R^2}\cosh^5(3\phi/4)\sinh(3\phi/4)e^{3\sqrt{2}\xi}, \\
\dot y=&-3{\rm sign}(h)y\sqrt{\frac{x^2}{2}-\frac{y^2}{2}+\frac{2}{9t_R^2}\cosh^6(3\phi/4) \exp\{3\sqrt{2}\xi\}}\\&
+\frac{2\sqrt{2}}{3t_R^2}\cosh^6(3\phi/4)e^{3\sqrt{2}\xi}.

\end{array}\right.\label{e:2sist}
\end{equation}
Notice that the potential (\ref{V-2new}) has an additional reflection symmetry
\begin{equation}
V_2(\xi,\phi) = V_2(\xi,-\phi).
\label{reflection}
\end{equation}
This provides the symmetry with respect to the origin in the plane $(\phi,x)$. The system (\ref{e:2sist})
has no stationary points. However, there is an interesting point
\begin{equation}
\phi=0,\  x = 0,
\label{interest}
\end{equation}
which freezes the dynamics of $\phi$ and hence, permits to consider independently the dynamics of $\xi$ and $y$, described by
the subsystem
\begin{equation}\left\{
\begin{array}{ll}
\dot\xi= &y, \\
\dot y=&-3{\rm sign}(h)y\sqrt{-\frac{y^2}{2}+\frac{2}{9t_R^2}e^{3\sqrt{2}\xi}}+\frac{2\sqrt{2}}{3t_R^2}e^{3\sqrt{2}\xi}.
\end{array}\right.\label{e:sis2dim}
\end{equation}
Apparently, the evolution of the universe is driven now by the phantom field and is subject to superacceleration.

In this case the qualitative analysis of the differential equations for $\xi$ and $y$, confirmed by the numerical simulations 
gives a predictable result: being determined by the only phantom scalar field the cosmological evolution is characterized by 
the growing positive value of $h$. Namely, the universe begins its evolution from the anti-Big Rip singularity ($ h = -\infty$)
then $h$ is growing  passing at some moment of time the value $h = 0$ (the point of minimal contraction of the cosmological radius $a(t)$) and then expands ending  its evolution in the Big Rip cosmological singularity ($h = +\infty$).

\begin{figure}[htp]\centering
\includegraphics{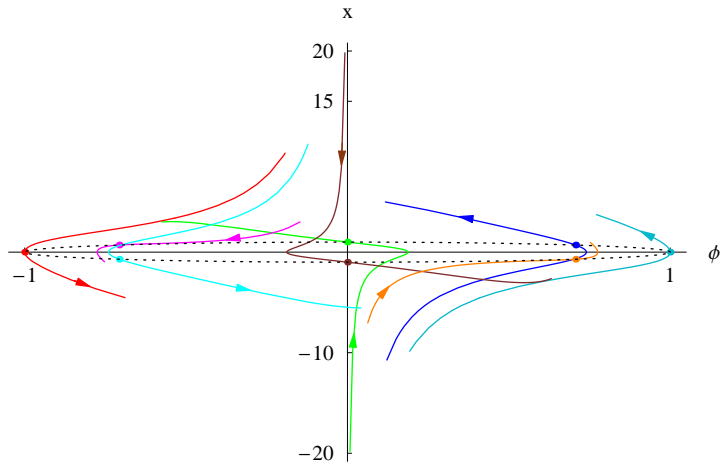}
\caption{An example of section of the 4D phase space for the model II obtained with numerical calculations. We can see a family of different trajectories corresponding to different choices of the initial conditions. Every initial condition in the graphic is chosen on the dashed ``ellipse''. The origin represents the point of freezing.}\label{figure3}
\end{figure}

Another numerical simulation can be done by fixing initial conditions for the phantom field as $\xi(0) = 0, y(0) = 0$ (see  Figure \ref{figure3}). 
Choosing various values of the initial conditions for the scalar field $\phi(0),x(0)$ around the point of freezing $\phi=0,
x=0$ we found two types of cosmological trajectories:\\
1. The trajectories starting from the anti-Big Rip singularity and ending in the Big Rip after the double crossing of the phantom 
divide line. These trajectories are similar to those discussed in the preceding subsection for the model I.\\
2. The evolutions of the type Bang-to-Rip. \\

Then we have carried out the numerical simulations of cosmological evolutions, choosing the initial conditions around 
the point of the phantomization point with the coordinates 
\begin{eqnarray}
&&\phi(0) =0, \nonumber \\
&&x(0) = \frac{\sqrt{2}}{3},\nonumber \\
&&\xi(0) = \frac43{\rm arctanh} \frac{1}{\sqrt{2}}, \nonumber \\
&&y(0) = \frac{\sqrt{2}}{3}.
\label{phant-point}
\end{eqnarray}

This analysis shows that in contrast to the model I, here the standard phantomization trajectory is stable and 
the trajectories of the type Bang-to-Rip are not exceptional, though less probable then those of the type 
anti-Big Rip to Big Rip. 

\section{Conclusion}
We have considered the problem of reconstruction of the potential in a theory with two scalar fields
(one standard and one phantom) starting with a given cosmological evolution. It is known ( see e.g. \cite{Star,Barrow,we-tach}) 
that in the
case of the only scalar field this potential is determined uniquely as well as the initial conditions for the scalar field,
reproducing the given cosmological evolution.  Changing the initial conditions, one can find a variety of 
cosmological evolutions, sometimes qualitatively different from the ``input'' one (see e.g. \cite{we-tach}).
In the case of two fields the procedure of reconstruction becomes much more involved. As we have shown here, there is a huge variety
of different potentials reproducing the given cosmological evolution (a very simple one in the case, which we have
explicitly studied here). Every potential entails  different cosmological evolutions, depending on the initial
conditions.


It is interesting that the existence of different dynamics of scalar fields corresponding to the same evolution of 
the Hubble parameter $h(t)$ can imply some observable consequences connected with the possible interactions of 
the scalar fields with other  fields. Indeed, in this case, the time dependence of the scalar fields considered above 
can directly affect physically observable quantities.
We are going to consider this topic elsewhere \cite{we-future}.


\section*{Acknowledgments}
We are grateful to G. Venturi for fruitful discussions and to M. Szydlowski, S.Yu. Vernov and A.I. Zhuk for useful correspondence.  This work was partially supported by Grants RFBR 08-02-00122-a, 05-02-17450  and  LSS-3251.2008.2, LSS-1757.2006.2.  The work of A.A. was also supported by  Grant FPA2007-66665, 2007PIV10046 and Program RNP 2.1.1.1112.


\section*{References}
\begin{thebibliography}{99}
\bibitem{cosmic}
Riess A et al. 1998
{\it Astron. J.} {\bf 116} 1009;
Perlmutter S J et al. 1999
{\it Astroph. J.} {\bf 517}, 565.
\bibitem{dark}
Sahni V and Starobinsky A A 2000 {\it Int. J. Mod. Phys.} D {\bf 9} 373;
Padmananbhan T 2003
{\it Phys. Rep.} {\bf 380} 235;
Peebles P J E and Ratra B 2003
{\it Rev. Mod. Phys.} {\bf 75} 559;
Sahni V 2002
{\it Class. Quantum Grav.} {\bf 19} 3435;
Copeland E J, Sami M and Tsujikawa S 2006
{\it Int. J. Mod. Phys.} D {\bf  15} 1753;
Sahni V and Starobinsky A A 2006 {\it Int. J. Mod. Phys.} D {\bf 15} 2105.
\bibitem{quintessence}
Caldwell R R, Dave R and Steinhardt P J 1998 
{\it Phys. Rev. Lett.} {\bf 80} 1582;
Frieman J A, Waga I 1998 {\it Phys. Rev.} D {\bf 57} 4642.
\bibitem{darkmodel}
Armendariz-Picon C, Damour T and Mukhanov V 1999 {\it Phys. Lett.} B
{\bf 458} 209; Armendariz-Picon C, Mukhanov V and Steinhardt P J 2000 {\it Phys. Rev. Lett.} {\bf 85} 4438; Armendariz-Picon C,
Mukhanov V, Steinhardt P J 2001 {\it Phys. Rev.} D {\bf 63} 103510;
Chiba T, Okabe T, Yamaguchi M 2000
{\it Phys. Rev.} D {\bf 62} 023511;
Kamenshchik A, Moschella U and Pasquier V 2001 {\it Phys. Lett.} B {\bf 511}
265;
Bilic N, Tupper G B, and Violeer R 2002 {\it Phys. Lett.} B {\bf 535} 17;
Fabris J C, Gonsalves V B and de Souza P E 2002 {\it Gen. Rel. Grav.}
{\bf 34} 53;
Bento M C, Bertolami O and Sen A A 2002 {\it Phys. Rev.} D {\bf 66} 043507;
Gorini V, Kamenshchik A and Moschella U 2003 {\it Phys. Rev.} D {\bf 67}
063509;
Sen A 2002 {\it JHEP} {\bf 0204} 048;
Gibbons G W 2002 {\it Phys. Lett.} B {\bf 537} 1; 
Fairbairn M and 
Tytgat M H G 2002 {\it Phys. Lett.} B {\bf 546} 1;
Mukohyama S 2002 {\it Phys. Rev.} D
{\bf 66} 024009;
Padmanabhan T and Choudhury T R 2002 {\it Phys. Rev.}
D {\bf 66} 081301(R);
 Sami M, Chingangbam P and Qureshi T 2002 {\it Phys. Rev.} D {\bf 66} 043530;
Felder G N, Kofman L and Starobinsky A A 2002 {\it JHEP} {\bf 0209} 026;
Frolov A, Kofman L and Starobinsky A 2002 {\it Phys. Lett.} B {\bf 545} 8;
Padmanabhan T 2002 {\it Phys. Rev.} D {\bf 66} 021301(R);
Feinstein A 2002 {\it Phys. Rev.} D {\bf 66} 063511;
Abramo L W and Finelli F 2003 {\it Phys. Lett.} B {\bf 575} 165;
Bagla J S, Jassal H K and Padmanabhan T 2003 {\it Phys. Rev.} D {\bf 67} 063504;
Gibbons G 2003 {\it Class. Quant. Grav.} {\bf 20} S321; Causse M B  A rolling tachyon field for both
dark energy and dark halos of galaxies {\it Preprint} astro-ph/0312206;
Paul B C and Sami M 2004 {\it Phys. Rev.} D {\bf 70} 027301;
Sen A 2005 {\it Phys. Scripta} T {\bf 117} 70;
Garousi M R, Sami M
and Tsujikawa S 2004 {\it Phys. Rev.} D {\bf 70} 043536;
Calcagni G 2004 {\it Phys. Rev.} D {\bf 69} 103508;
Aguirregabiria J M and Lazkoz R 2004 {\it Mod. Phys. Lett.} A {\bf 19} 927;
Aguirregabiria J M and Lazkoz R 2004 {\it Phys. Rev.} D {\bf 69} 123502;
Herrera R, Pavon D and
Zimdahl W 2004 {\it Gen. Rel. Grav.} {\bf 36} 2161;
Barnaby N 2004 {\it JHEP}
{\bf 0407} 025;
Allemandi G, Borowiec A and Francaviglia M 2004 {\it Phys. Rev.} D {\bf 70} 043524;
Zimdahl W  Interacting Dark Energy and Cosmological Equations of State {\it Preprint} gr-qc/0505056;
Allemandi G, Borowiec A, Francaviglia M and Odintsov S D 2005
{\it Phys. Rev.} D {\bf 72} 063505.
\bibitem{before}
Cooper F and Venturi G 1981
{\it Phys. Rev.} D {\bf 24} 3338;
Ratra B and Peebles P J E 1988 {\it Phys. Rev.} D {\bf 37} 3406;
Peebles P J E and Ratra B 1988 {\it Astrophys. J. Lett.} {\bf 325} L17;
Wetterich C 1988 {\it Nucl. Phys.} B {\bf 302} 668.
\bibitem{phantom}
Caldwell R R 2002 {\it Phys. Lett.} B {\bf 545} 23;
Gonzalez-Diaz P F
{\it Phys. Lett.} B {\bf 586} 1;
Singh P, Sami M and Dadhich N {\it Phys.  Rev.}  D {\bf 68} 023522;
Nojiri S and Odintsov S D {\it Phys. Lett.} B {\bf 562} 147;
Capozziello S, Nojiri S and Odintsov S D
{\it Phys. Lett.} B {\bf 632} 597;
Johri V B 2004
{\it Phys. Rev.}  D {\bf 70} 041303(R);
Onemli V K and Woodard R P 2000 {\it Class. Quant. Grav.} {\bf 19} 4607;
Hannestad S and Mortsell E 2002 {\it Phys. Rev.} D {\bf 66} 063508;
Carroll S M, Hoffman M and Trodden M
{\it Phys.  Rev.}  D {\bf 68} 023509;
Frampton P H  Stability Issues for $w < -1$ Dark Energy {\it Preprint} hep-th/0302007;
Elizalde E, Nojiri S and Odintsov S D
{\it Phys. Rev.} D {\bf 70} 043539;
Gonzalez-Diaz P F and Siguenza C L 2004
{\it Nucl. Phys.} B {\bf 697} 363;
Gibbons G W Phantom Matter and the Cosmological Constant {\it Preprint} hep-th/0302199;
McInnes B  What If $w < -1$ ? {\it Preprint} astro-ph/0210321;
Chimento L P and Lazkoz R 2003 {\it Phys. Rev.  Lett.} {\bf 91} 211301;
Dabrowski M P, Stachowiak T and Szydlowski M 2003
{\it Phys. Rev.} D {\bf 68} 103519;
Singh P, Sami M and Dadhich N 2003
{\it Phys. Rev.} D {\bf 68} 023522;
Gonzalez-Diaz P F 2004 {\it Phys. Rev.}  D {\bf 69} 063522;
Onemli V K and Woodard R P 2004
{\it Phys. Rev.} D {\bf 70} 107301;
Sami M and Toporensky A 2004
{\it Mod. Phys. Lett.} A {\bf 19} 1509;
Stefancic H 2004 {\it Eur. Phys. J. C} {\bf 36} 523;
Stefancic H 2004 {\it Phys. Lett.} B {\bf 586} 5;
Brunier T, Onemli V K and  Woodard R P 2005
{\it Class. Quantum  Grav.}  {\bf 22} 59;
Santos J and Alcaniz J S 2005
{\it Phys. Lett.} B {\bf 619} 11;
Carvalho F C and Saa A 2004 {\it Phys. Rev.} D {\bf 70} 087302;
Melchiorri A, Mersini L,  Odman C J and Trodden M 2003
{\it Phys. Rev.} D {\bf 68} 043509;
Cline J M, Jeon S and Moore J D 2004
{\it Phys. Rev.} D {\bf 70} 043543;
Guo Z K, Piao Y S, Zhang X M and  Zhang Y Z 2005
{\it Phys. Lett.} B {\bf 608} 177;
Feng B, Wang  X L and Zhang X M 2005
{\it Phys. Lett.} B {\bf 607} 35;
Aref'eva I Y, Koshelev A S and  Vernov S Y Exactly Solvable SFT Inspired Phantom Model to be published in {\it Theor. Math. Phys.} {\it Preprint} astro-ph/0412619;
Zhang X F, Li H, Piao Y S and  Zhang X M 
Two-field Models of Dark Energy with Equation of State Across -1 {\it Preprint} astro-ph/0501652;
Calcagni G 2005 {\it Phys. Rev.} D {\bf 71} 023511;
Li M, Feng B and Zhang X 2005 {\it JCAP} {\bf 0512} 002;
Babichev E, Dokuchaev V and  Eroshenko Yu 2005
{\it Class. Quant. Grav.} {\bf 22} 143;
Anisimov A, Babichev E and Vikman A 2005
{\it JCAP} {\bf 0506} 006;
Nojiri S, Odintsov S D and Tsujikawa S 2005
{\it Phys. Rev.} D {\bf 71} 063004;
Gumjudpai B, Naskar T,  Sami M and Tsujikawa S 2005 {\it JCAP} {\bf 0506} 007;
Sami M, Toporensky A, Tretjakov P V and Tsujikawa S 2005 {\it Phys. Lett.} B
{\bf 619} 193;
Nojiri S and Odintsov S D 2006 {\it Gen. Relativ. Grav.} {\bf 38} 1285;
Brevik I 2006 {\it Gen. Relativ. Grav.} {\bf 38} 1317.
\bibitem{phant-obs}
Alam U, Sahni V, Saini T D and Starobinsky A A
2004 {\it Mon. Not. Roy. Astron. Soc.} {\bf 354} 275;
Padmanabhan T and Choudhury T R 2003
{\it Mon. Not. Roy. Astron. Soc.} {\bf 344} 823;
Choudhury T R and  Padmanabhan T 2005 {\it Astron.Astrophys.} {\bf 429} 807;   
Wang Y and Mukherjee P 2004 {\it Astrophys. J.}   {\bf 606} 654;
Huterer D and Cooray A 2005 {\it Phys. Rev.} D {\bf 71} 023506;
Daly R A and  Djorgovski S G 2003 {\it Astrophys.  J.} {\bf 597} 9;
Alcaniz J S 2004 {\it Phys. Rev.} D {\bf 69} 083521;
 Lima J A S,  Cunha J V and Alcaniz J S 2003 {\it Phys. Rev.} D {\bf 68} 023510;
Alam U, Sahni V and Starobinsky A A 2004  {\it JCAP} {\bf 0406} 008;
Wang Y and Freese K 2006 {\it Phys. Lett.} B {\bf 632}449;
Upadhye A, Ishak M and Steinhardt P J 2005 {\it Phys. Rev.} D {\bf 72} 063501;
Dicus D A and Repko W W 2004 {\it Phys. Rev.}  D {\bf 70} 083527;
Espana-Bonet C and Ruiz-Lapuente P Dark Energy as an Inverse Problem
{\it Preprint} hep-ph/0503210;
 Wei Y H The power-law expansion universe and dark energy evolution
 {\it Preprint} astro-ph/0405368;
Jassal H K, Bagla J S, and Padmanabhan T 2005 {\it Mon. Not. Roy. Astron. Soc. Letters} {\bf 356} L11;
Nesseris S and Perivolaropoulos L 2004 {\it Phys. Rev.} D {\bf 70} 043531;
Lazkoz R, Nesseris S and Perivolaropoulos L 2005 {\it JCAP} {\bf 0511} 010;
Jassal H K,  Bagla J S and Padmanabhan T 2005 {\it Phys. Rev.} D {\bf 72} 103503;
Feng B, Li M, Piao Y S  and Zhang X 2006 {\it Phys. Lett.} B {\bf 634} 101-105;
Xia J Q, Feng B and Zhang X M 2005 {\it Mod. Phys. Lett.} A {\bf 20} 2409.
\bibitem{phant-obs1}
Cabre A, Gaztanaga E, Manera M, Fosalba P and Castander F 2006 {\it Mon. Not. Roy. Astron. Soc. Lett.} {\bf 372} L23-L27.
\bibitem{obs-new}
 Barger V, Gao Y and Marfatia D 2007 {\it Phys. Lett.} B {\bf 648} 127;
Gong A and Wang A 2007 {\it Phys. Rev.} D {\bf 75} 0435520;
Alam U, Sahni V and Starobinsky A A 2007 {\it JCAP} {\bf 0702} 011;
Nesseris S and Perivolaropoulos L 2007 {\it JCAP} {\bf 0702} 025;
Zhao G B, Xia J Q, Li H et al. Probing for dynamics of dark energy and curvature of universe with latest cosmological observations {\it Preprint} astro-ph/0612728;
Serra P, Heavens A and Melchiorri A 2007 {\it MNRAS} {\bf 379} 1,169;
Davis T M, Mortsell E, Sollerman J et al. Scrutinizing Exotic Cosmological Models Using ESSENCE Supernova Data Combined with Other Cosmological Probes {\it Preprint} astro-ph/0701510;
Wright E L Constraints on Dark Energy from Supernovae, Gamma Ray Bursts, Acoustic Oscillations, Nucleosynthesis and Large Scale Structure and the Hubble constant {\it Preprint} astro-ph/0701584;
Wang Y and Mukherjee P Observational Constraints on Dark Energy and Cosmic Curvature {\it Preprint} astro-ph/0703780.
\bibitem{divide}
Boisseau B, Esposito-Farese G, Polarski D and Starobinsky A A
2000 {\it Phys. Rev. Lett.} {\bf 85} 2236;
Esposito-Farese G and Polarski D 2001 {\it Phys. Rev.} D {\bf 63} 063504;
Vikman A 2005
{\it Phys.  Rev.}  D {\bf 71} 023515;
Perivolaropoulos L 2005
{\it Phys.  Rev.} D {\bf 71} 063503;
McInnes B 2005 {\it Nucl. Phys.} B {\bf 718} 55;
Aref'eva I Ya, Koshelev A S and Vernov S Yu 2005  {\it Phys. Rev.} D {\bf 72} 064017;
Perivolaroupoulos L 2005  {\it JCAP} {\bf 0510} 001;
Caldwell R R and Doran M 2005 {\it
Phys. Rev.} D {\bf 72} 043527;
Sahni V and Shtanov Yu 2003 {\it JCAP} {\bf 0311} 014;
Sahni V and Wang L 2000 {\it Phys. Rev.} D {\bf 62} 103517;
Wei H, Cai R G and Zeng D F 2005 {\it Class. Quant. Grav.} {\bf 22} 3189;
Wei H, Cai R G 2005 {\it Phys. Rev.} D {\bf 72} 123507;
Wei H, Cai R G 2006 {\it Phys. Lett.} B {\bf 634} 9;
Cai R G and Wang A 2005 {\it JCAP} {\bf 0503} 002;
Gannouji R, Polarski D, Ranquet A and Starobinsky A A 2006 {\it JCAP} {\bf 0609} 016.
\bibitem{we}
Andrianov A A, Cannata F and Kamenshchik A Y 2005 {\it Phys. Rev.} D {\bf 72} 043531.
\bibitem{we1}
Cannata F and Kamenshchik A Yu  Networks of cosmological histories, crossing of the phantom
divide line and potentials with cusps (to appear in {\it Int. J. Mod. Phys.} D) {\it Preprint} gr-qc/0603129.
\bibitem{two-field}
Andrianov A A, Cannata F and Kamenshchik A Yu 2006 {\it Int. J. Mod. Phys.} D {\bf 15} 1299;
Andrianov A A, Cannata F and Kamenshchik A Yu 2006 {\it J. Phys.} A {\bf 39} 9975.
\bibitem{two-field1}
Feng B, Wang X and Zhang X 2005 {\it Phys. Lett.} B {\bf 607} 35;
Guo Z K, Piao Y S, Zhang X and Zhang Y Z 2005 {\it Phys. Lett.} B {\bf 608} 177;
Hu W 2005 {\it Phys. Rev.} D {\bf 71} 047301;
Perivolaropoulos L 2005 {\it Phys. Rev.} D {\bf 71} 063503;
Caldwell R R and Doran M 2005 {\it Phys. Rev.} D {\bf 72} 043527.
\bibitem{Star}
Starobinsky A A 1998 {\it JETP Lett.} {\bf 68} 757.
\bibitem{Barrow}
Burd A B and Barrow J D 1988 {\it Nucl. Phys.} B {\bf 308} 929;
Barrow J D 1990 {\it Phys. Lett.} B {\bf 235} 40.
\bibitem{we-tach}
Gorini V, Kamenshchik A, Moschella U and Pasquier V 2004 {\it Phys. Rev.} D {\bf 69} 123512.
\bibitem{Yurov}
Zhuravlev V M, Chervon S V and Shchigolev V K 1998 {\it JETP} {\bf 87} 223;
Chervon S V and Zhuravlev V M The cosmological model with an analytic exit from inflation {\it Preprint} gr-qc/9907051;
Yurov A V Phantom scalar fields result in inflation rather than Big Rip {\it Preprint} astro-ph/0305019;
Yurov A V and Vereshchagin S D 2004 {\it Theor. Math. Phys.} {\bf 139} 787.
\bibitem{Ohta}
Guo Z K, Ohta N and Zhang Y Z 2007
 {\it Mod. Phys. Lett.}  A {\bf 22} 883; 
Guo Z K, Ohta N and Zhang Y Z 2005
 {\it Phys.\ Rev.}  D {\bf 72} 023504. 
\bibitem{Zhuk}
Zhuk A 1996  {\it Class. Quant. Grav.} {\bf 13} 2163.
\bibitem{Szydlowski}
Szydlowski M and Czaja W 2004
{\it Phys. Rev.} D {\bf 69} 083507;
Szydlowski M and Czaja W 2004
{\it Phys. Rev.} D {\bf 69} 083518; 
Szydlowski M 2005 {\it Int. J. Mod. Phys.} A {\bf 20} 2443;
Szydlowski M, Hrycyna O and  Krawiec A 2007 {\it JCAP} {\bf 0706} 010.
\bibitem{Vernov}
Vernov S Yu  Construction of Exact Solutions in Two-Fields Models and
the Crossing of the Cosmological Constant Barrier {\it Preprint} astro-ph/0612487.
\bibitem{Regoli}
Regoli D 2007 {\it Mg. Thesis} University of Bologna (in Italian).
\bibitem{Bel-Khal}
Belinsky V A, Khalatnikov I M, Grishchuk L P and Zeldovich Ya B 1985 {\it Sov. Phys. JETP} {\bf 62} 195;
Belinsky V A, Khalatnikov I M, Grishchuk L P and Zeldovich Ya B 1985 {\it Phys. Lett.} B {\bf 155} 232.
\bibitem{we-future}
Andrianov A A, Cannata F, Kamenshchik A Yu and Regoli D, work in progress.
\end{thebibliography}
\end{document}